\documentstyle[epsfig]{aipproc}

\newcommand{\be}{\begin{equation}}
\newcommand{\ee}{\end{equation}}
\newcommand{\ba}{\begin{eqnarray}}
\newcommand{\ea}{\end{eqnarray}}
\newcommand{\dle}[1]{\label{#1}}

\newcommand{\dr}[1]{\ref{#1}}
\newcommand{\dc}[1]{\cite{#1}}

\newcommand{\veca}{{\mathbf{a}}}
\newcommand{\vecb}{{\mathbf{b}}}
\newcommand{\vecx}{{\mathbf{x}}}

\newcommand{\si}{{\sigma}}
\newcommand\spr[1]{\mathaccent19{#1}}

\begin{document}
\title{Strings after D-term inflation: evolution and properties
of chiral cosmic strings}

\author{D.A. Steer}
\address{D\'epartement de Physique Th\'eorique, 24 Quai Ernest Ansermet,
Universit\'e de Gen\`eve, \\
1211 Gen\`eve 4, Switzerland 
}

\maketitle

\begin{abstract} We motivate the study of chiral cosmic strings through a 
scenario of structure formation which mixes D-term inflation and strings.  
We then discuss some properties of chiral
cosmic strings, and results regarding their evolution and
possible cosmological consequences are presented.
\end{abstract}

\section*{Introduction and Motivation}

A scenario of structure formation which has attracted some
attention recently, and which was mentioned a number of times during this 
conference,  is that of `inflation and strings'.  In these
mixed models, a symmetry is broken at the end of inflation
leading to the formation cosmic strings whose properties (whether 
they are current carrying, local or global etc.) will depend on
the details of the model.
In the literature, 
predictions for the temperature fluctuations in the CMB have been made for 
various combinations of inflation with local Nambu-Goto (NG) strings or global 
strings in different cosmologies \cite{Carlo}.  
There the $C_l$'s were decomposed as
\be
C_l = \alpha C_l^{{\rm inflation}} + (1-\alpha)C_l^{\rm strings} 
\qquad (0 \leq \alpha \leq 1),
\dle{cl}
\ee
and an extensive literature exists on 
$C_l^{\rm strings}$ from NG and global strings.

Whilst it is interesting to `add' the effects of 
strings and inflation in this way, it is important 
specify which theories actually produce such a combination
so as to understand exactly the properties of the strings generated.  
One motivation for studying chiral cosmic strings
comes from the 
well known SUSY D-term inflation model \dc{Rachel} which was 
considered  in the context of structure formation in \dc{CHM}.  
There equation (\dr{cl}) was used, 
assuming that
the strings formed are NG strings.
Whilst there may be models in which NG strings are formed at 
the end of inflation, it appears to be less well known that the strings
formed at the end of D-term inflation
are not NG 
but {\em chiral} cosmic 
strings \cite{DDT}.  Hence to obtain $C_l$ using (\dr{cl}) one 
should calculate $C_l^{\rm chiral \; strings}$: this has not so far 
been done.  Here we comment some recent progress
made in this direction \dc{CP,US,THEM}.

Chiral cosmic strings are current carrying strings for which
the world sheet current $j^i$ ($i=0,1$; the 2D world sheet 
metric raises and lowers indices) is null:
\be
j_i j^i = j^2 = 0
\dle{j2}.
\ee
Generically, strings may well have an internal structure and
carry a current (which could also be space- or time-like).  For
example, suppose the Higgs field $\Phi$ which forms the string couples to
other fields in the theory, say fermionic fields $\Psi$ via a
Yukawa term; ${\cal{L}} = \ldots - g \bar{\Psi}_L \Psi_{R}
\Phi - g
\bar{\Psi}_R \Psi_{L} \Phi^*$.  (Such a coupling might well be expected!)  
Then since $\Phi$ vanishes inside the core of the string,
massless modes exist there.  The string is chiral if these modes
propagate in only one direction along the string core: this indeed 
happens for the strings formed at the end of D-term inflation \dc{DDT}.

Just as for NG strings, $C_l^{\rm chiral \; strings}$ can only
be calculated once 1)  the dynamics and properties of
individual chiral strings, and interactions between them are
understood, and 2) the evolution of the chiral string network has been 
determined.

Before commenting on some aspects of 1) and 2) below, recall that
chiral and NG strings share some common properties.
In particular, the metric around chiral and NG strings 
is the same so that individual strings of each kind
generate the same temperature difference in traveling photons.  
However, the evolution of networks of chiral and NG strings is probably 
very different (see below):  depending on magnitude 
of the current on the chiral strings, these may never self-intersect so 
that the network can be `frozen' producing a cosmological catastrophe.
Throughout we work in Minkowski space with metric $\eta_{\mu \nu}
= (+,-,-,-)$.

\section*{Chiral cosmic strings}

\paragraph*{Action and equations of motion:}

The effective 2D chiral string action has two terms: the
usual NG action plus a second term resulting from the zero modes
moving along the string \dc{CP}.  Let $\phi$ be a dimensionless real
scalar field (the phase of the charge carriers) defined on the
world sheet which is labeled by coordinates $\sigma^{i}$. Then the action, 
first proposed by Carter and Peter \dc{CP}, is
\begin{equation}
S = - \int d^2 \sigma\, \sqrt{-\gamma} \left(m^2 - \frac{1}{2}
\psi^2 \gamma^{ij} \phi_{,i} \phi_{,j}\right)
\dle{Baction}
\end{equation}
where $\gamma_{ij} = \eta_{\mu \nu}x^{\mu}_{,i} x^{\nu}_{,j}$ is
the induced world sheet metric
and $x^{\mu}(\sigma^0,\sigma^1)$ is the position of the
string.  The dimensionless Lagrange multiplier $\psi^2$ sets the
constraint given in equation (\dr{j2}) namely that the 
current $j^i = \psi \phi^{,i}$ is null.
The corresponding conserved 
charge       
 is $C = \int d\sigma^i \epsilon_{ik}j^{k}$
where $\epsilon$ is the antisymmetric surface measure tensor whose
square gives the induced metric; $\gamma_{ij} =
\epsilon_{ik}\epsilon^{k}_{ \; j}$.\footnote{$C$ is closely related [8]
to usual charges $N$ and $Z$ discussed for example in [9]. }

As was first shown in \dc{CP} and later, in perhaps more familiar notation to
cosmic string physicists in \dc{US,THEM}, the equations of motion obtained
from (\dr{Baction}) take a very simple and familiar integrable form 
{\em provided} one makes a careful gauge choice: they are
\be
\ddot{\vecx} - \vecx'' = 0 \qquad
\Longrightarrow \qquad \vecx(t,\si) = [\veca(t+\si) +
\vecb(t-\si)]/2
\dle{eqn}
\ee
where $\dot{\vecx} = 
\partial \vecx /\partial t$ with $t$ background time; and $\vecx' = \partial \vecx/\partial \sigma $ where $\sigma$ measures the energy per unit
length; $E =m^2 \int d\sigma$.  
The vectors $\veca$ and $\vecb$ must satisfy the constraints
\be
\spr{\veca}^2 =1 \qquad , \qquad 
k^2 := \spr{\vecb}^2  \le 1
\dle{con}
\ee
where for instance $\spr\veca(q)\equiv d\veca(q)/dq.$   The magnitude of $k$
determines the conserved charge on the string which is 
given\footnote{This expression 
corrects a factor of 2 error in \dc{US}: see
\dc{ME} for details.} by
$C = \frac{m}{2} \int d\sigma  \sqrt{1-k(\sigma)^2}$.  
Hence in the NG limit for which 
$k=1$, the charge vanishes as required.

\paragraph*{Some general properties:}

Note first that from the constraints (\dr{con}), ${\vecx'} \neq 0$ and 
$|\dot{\vecx}| \neq 1$.  Thus chiral cosmic strings cannot have cusps.

Secondly, observe that for any general {\em arbitrarily} shaped 
string, the limit $k=0$ (which corresponds to maximal charge)
is special since here $\dot{\vecx} = \vecx'$ with
$|\dot{\vecx}| = |\vecx'| = 1/2$.  Thus the string moves along itself at
half the speed of light and
appears to be
stationary (as the only visible component of velocity is that perpendicular
to the string): it cannot 
self-intersect.  If the strings form a loop these 
non-self-intersecting solutions are called vortons and they
can be of {\em any} shape, though their length should
be $\gg m^{-1}$ as otherwise they would decay immediately into particles
\dc{CD}.  
Thus if the state $k=0$ were reached for each string either at the time
of formation of the network or at any later time, 
the network would remain frozen and dominate the energy 
density of the universe.   
In the D-term inflation model, the initial 
charge might indeed be expected to be close to maximal following the 
arguments given in \dc{CD}.  This point will be discussed in detail in
 \dc{ACMP}.  However, it is not clear that
the charge should be independent of the position on the string (i.e.\ why
$k(\sigma-t)$ should be constant), especially 
for strings larger than the horizon or for loops that are formed 
as the result of
self-intersections.

Finally, observe that $\dot{\vecx} \neq 0$ and that there is always a component
of velocity along the string in this gauge.  This suggests that a loop of 
string has angular momentum, which is indeed the case and is 
responsible for the fact that the above mentioned vorton solutions can exist.

\paragraph*{Loop self-intersections:}

As for NG strings, it is essential that energy should 
be lost from the chiral string network 
as if not the strings would dominate the
energy density in the universe.  
We have seen that this problem occurs if $k=0$.
If $k \neq 0$ energy loss might be 
expected to occur, as for NG strings, through loop-self intersections and decay
into gravitational radiation.  For this to happen, the loops must
self-intersect:  we have studied the self-intersection 
probability of chiral string loops as a function of $k$ and 
the `wiggliness' of the string measured in terms of the number of harmonics
in $\veca$ and $\vecb$ (these must be periodic for loops).  
The results are shown in figure 1 for constant \dc{US} and non-constant
\dc{ME} $k$.
\begin{figure}[ht]
\centerline{\hbox{
\psfig{figure=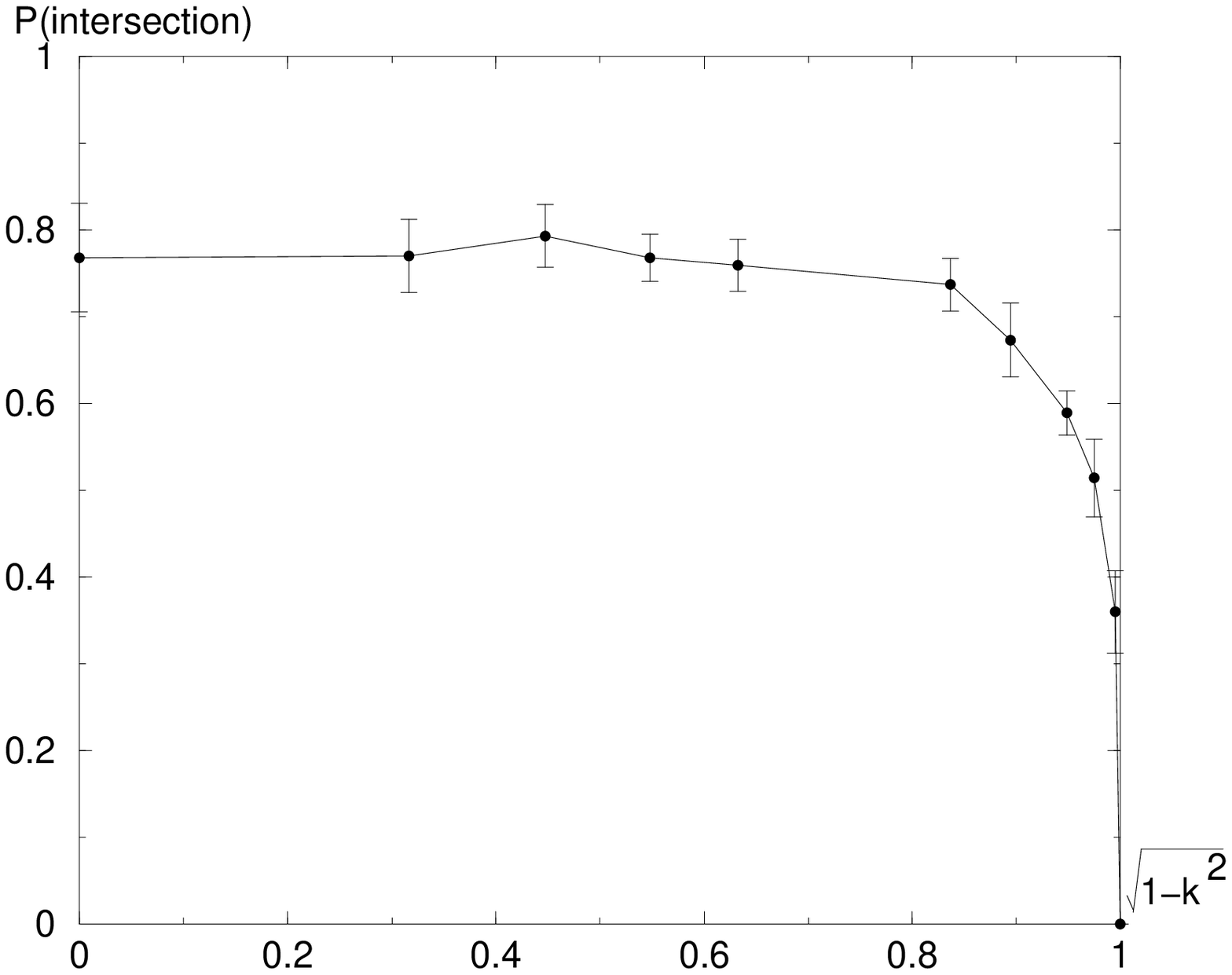,height=6.0cm,width=6.5cm}
\psfig{figure=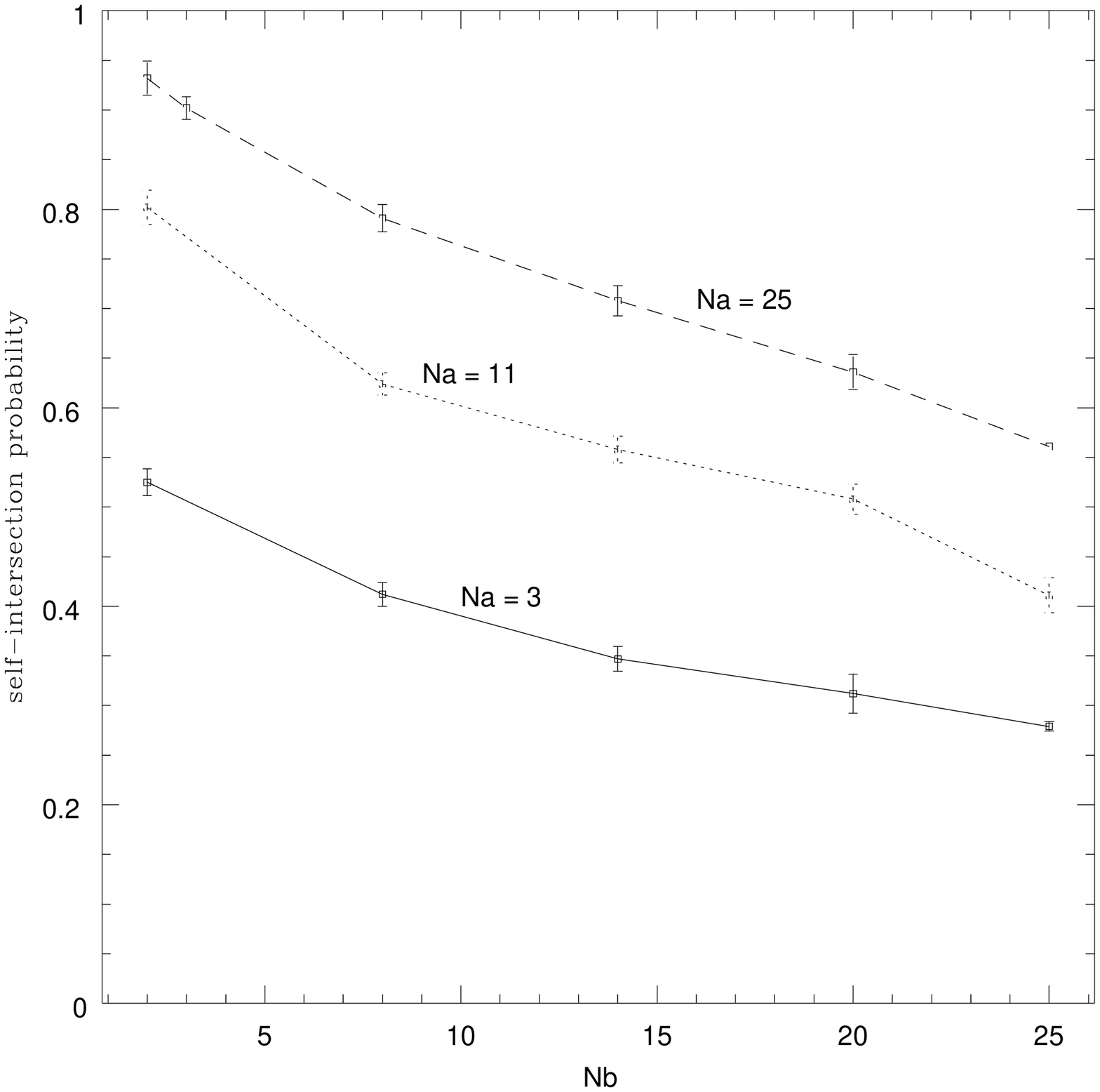,height=6.5cm,width=6.5cm}}}
\caption{Loop self-intersection probability $P_I$.  Lefthandside: 
$k=$constant and $\veca$ and $\vecb$ both have 5 
harmonics (Na=Nb=5) [6].
Righthandside: Non-constant $k(q)$ which vanishes at 2(Nb) points 
only (see [8] for details): $P_I$ decreases with increasing Nb.
All curves have same $C$. }
\label{Rcbig}
\end{figure}

\section*{Conclusions}

The starting point for this study of chiral strings was the
well-defined, unique action first proposed in \dc{CP}.  With a
suitable gauge choice the resulting equations of motion are 
integrable \dc{CP,US,THEM}
(see equation (\dr{eqn})) and the solutions are parametrised by the
function $k(\sigma -t)$ which determines the total charge on the string.
We saw that if $k=0$ the strings do not move and never self-intersect
leading to a cosmological catastrophe.  If $k(q) \neq 0$,
the loops can in some cases 
self-intersect, though the ultimate fate of the daughter 
loops still remains to be understood.

\paragraph*{Acknowlegements}

I would like to thank my collaborators 
Anne Davis, Tom Kibble and Mike Pickles;
Ola
T\"ornkvist for a useful discussion; and
PPARC (UK) and an ESF network for financial support.

\end{document}